\documentclass[twocolumn,showpacs,aps,floatfix]{revtex4}

\usepackage{amsmath}
\usepackage{graphicx}
\usepackage{dcolumn}
\usepackage{bm}

\begin{document}
\preprint{ND Atomic Experiment 2003-2}
\title{Beam-gas Spectroscopy of Sextet Transitions in O$^{3+}$, F$^{4+}$ and Ne$^{5+}$}
 \author{Bin Lin$^{1}$}
 \email{blin@nd.edu}
 \homepage{http://www.nd.edu/~blin/}
 \author{H. Gordon Berry$^{1}$}
 \email{Berry.20@nd.edu}
 \homepage{http://www.science.nd.edu/physics/Faculty/berry.html}
 \author{Tomohiro Shibata$^{1}$}
  \author{Lanlan Lin$^{2}$}
\affiliation{1 Department of Physics, University of Notre Dame, Notre Dame, IN 46556\\
2 Ames Lab, Ames, IA 50515}
\date{\today}
\begin{abstract}
We present VUV observations of transitions between doubly excited
sextet states in O$^{3+}$, F$^{4+}$ and Ne$^{5+}$. Spectra were
produced by collisions of an oxygen, fluorine and neon beam with a
nitrogen gas jet target. Prepared beam-gas experiment yields new
and explicit information on doubly core-excited ions. Some
observed lines were assigned to the 1s2s2p$^{3}$
$^{6}$S-1s2p$^{3}$3s, 3d $^{6}$P electric-dipole transitions in
O$^{3+}$, F$^{4+}$ and Ne$^{5+}$. Three lines have been
reassigned. Present data are the first explicit measurements on
transitions between sextet states in boronlike ions by beam-gas
spectroscopy.
\end{abstract}

 \maketitle

The sextet states of doubly excited boronlike ions are possible
candidates for x-ray and VUV-lasers~\cite{lin}, and have been
investigated
recently. The lowest terms of this system (1s2s2p$^{3}$ $^{6}$%
S, 1s2s2p$^{2}$3s $^{6}$P and 1s2s2p$^{3}$3d $^{6}$P) have been
studied along the B I sequence~\cite{bl,lap,mie}. The studies of
higher excited sextet states (1s2p$^{3}$3s $^{6}$S) have been
lately reported~\cite{lin}. However, the energy level diagrams of
these ions are still far from complete. Experimentally, these
levels are difficult to observe by conventional spectroscopy
techniques, such as the high voltage discharge in gas cell method,
because they lie well above several ionization limits of
five-electron singlet states. Even though they are metastable
against autoionization, they usually de-excite and disappear by
collisions with other ions without radiative transitions. The fast
beam-foil technique allows straight forward observations of the
radiative transitions produced by these sextet
states~\cite{berry1,kla} along B I sequence.

In 1992 the beam-foil spectroscopy~\cite{bl,lap} was used to
provide initial data on low-lying sextet states in boronlike
nitrogen, oxygen and fluorine. The recent work of Lapierre and
Knystautas~\cite{lap} on possible sextet transitions in Ne VI
highlights the significance in this sequence. They measured
several excitation energies and lifetimes. The fine structures of
individual 1s2s2p$^{2}$3s $^{6}$P$_{J}$ states were resolved and
measured in O IV, F V and Ne VI by Lin and Berry et al~\cite{lin}.
There are no further results reported for transitions from sextet
states in boronlike system.

In some works on beam-foil spectroscopy of sextet states in B I
isoelectronic sequence, their identifications show rather weak
lines and overwhelming blending problems. Hence, accurate
theoretical studies of sextet states in doubly excited B I
isoelectronic sequence are strongly needed to help
identifications. However, the theoretical analysis of these
five-electron ions is difficult because strong electron
correlation, relativistic corrections, and even QED effects have
to be included in the
calculations~\cite{mie,lin,kt1,fibk,MCDF,MCDF1,MCDF2,qed1,qed2,drake}.

Doublet and quartet transitions in core-excited boronlike ions
have been studied successfully using beam-gas collision
spectroscopy~\cite{des1,des2}. Well prepared beam-gas spectroscopy
of oxygen, fluorine and neon ions is a possible technique to study
the sextet transitions in boronlike O IV, F V and Ne VI. The
sextet states in B I isoelectronic sequence are well above several
ionization levels and metastable against electric-dipole radiation
decay to singly excited five-electron states and against Coulomb
autoionization into the adjacent continuum 1s$^{2}$2l'2l''nl
$^{4}$L due to different spin multiplicity. Thus, the main decay
channel is radiation in beam-gas experiments. The lifetimes of the
ground quintet states in boronlike ions are long enough (about
10$^{-5}$ second for O V) so that the "prepared" beryllium-like
ions, excited to metastable quintet states by collisions with
nitrogen gas molecular, can reach alkaline vapor target atoms. The
spectra obtained have significant features emitted from
core-excited ions and very clean background because of, after
careful preparation and selection of the ion beam, the dominant
core-conserving single-electron pick-up cross-sections of
metastable doubly excited ions at low beam energy in alkaline
vapor cell.

Presented herein are the first measurements on the
1s2s2p$^{3}$ $^{6}$S-1s2s2p$^{2}$3s, 3d $%
^{6}$P electric-dipole transitions in boronlike  O IV, F V and Ne
VI by well prepared beam-gas spectroscopy. Comparison is given
with the recent results of measurements from beam-foil
spectroscopy~\cite{bl,lap}, and multi-configuration Hartree-Fock
(MCHF) and multi-configuration Dirac-Fock (MCDF) calculations
~\cite{bl,mie,lap} for the transitions.

\begin{figure}[tbp]
\centerline{\includegraphics*[scale=0.69]{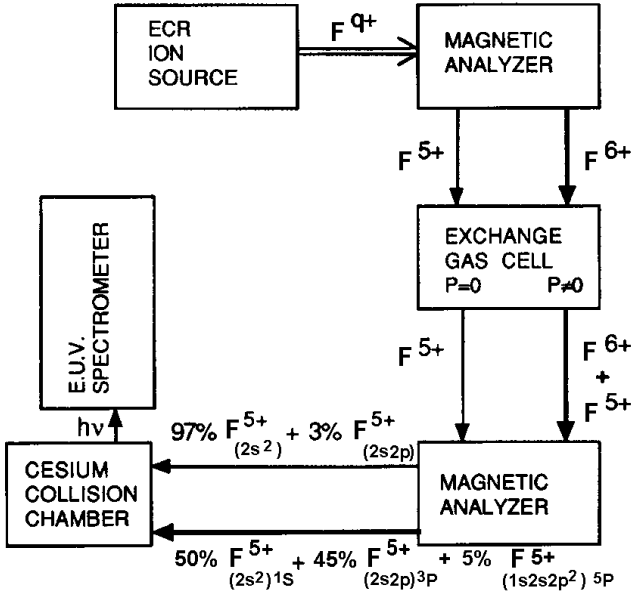}} \caption{Block
diagram of the beam-gas experimental apparatus showing the
procedures to be followed to prepare F$^{5+}$ beam in low (3\%)
and high (48\%) fractions of the metastable states.} \label{Fig1}
\end{figure}

\begin{figure}[tbp]
\centerline{\includegraphics*[scale=0.60]{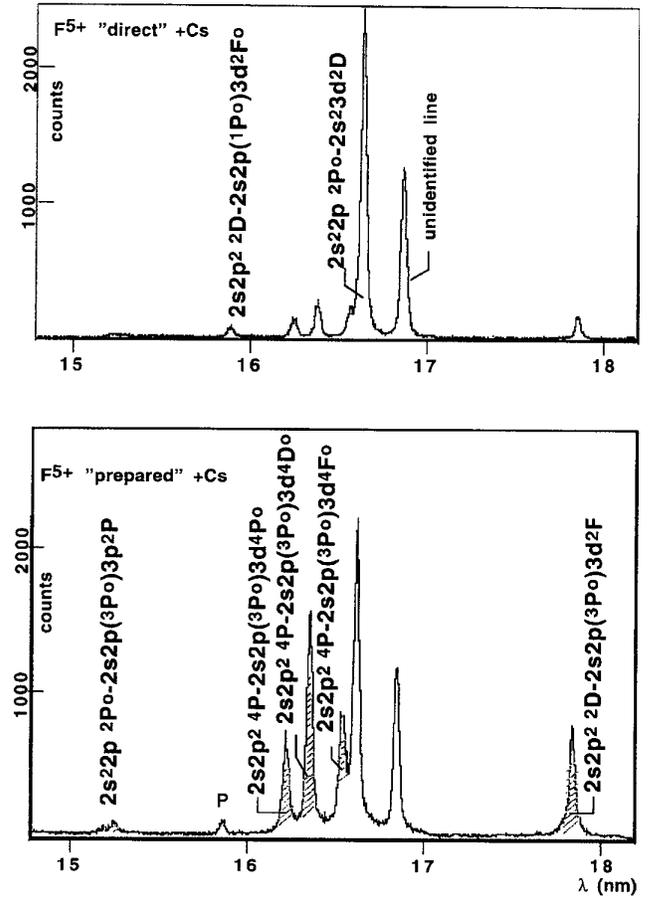}}
\caption{Spectra of F V in the 150-180 \AA \ region for direct and
prepared beams. The transition labelled as P is the 1s2s2p$^{3}$ $^{6}$S%
 - 1s2s2p$^{2}$3s $^{6}$P electric-dipole transitions in F V.} \label{Fig2}
\end{figure}

In this work, the beam-gas spectra of oxygen, fluorine and neon at
low beam energy were previously recorded at Lyon using grating
incidence spectrometers. The experimental arrangement in Fig. 1
was used~\cite{des1,des2}. Multi-charged ions extracted from a 14
MHz CAPRICE ECR ion source of the AIM, a joint CEA/CNRS facility
at CEA Grenoble with acceleration voltage of 2-20 kV, were mass
and charge analyzed by two bending magnets and sent into the beam
line devoted to UV spectroscopy. A nitrogen gas jet target was
placed between the two magnets. After electron capture collisions
in a cesium cell, photons emitted were detected at 90 degrees to
the ion beam direction by a 2.2 m- McPherson grazing incidence
spectrometer equipped with a position-sensitive microchannel plate
detector which allows simultaneous recording of spectral lines
within a wavelength region of about 50 \AA \ in the range of
60-600 \AA .

\begin{figure}[tbp]
\centerline{\includegraphics*[scale=0.85]{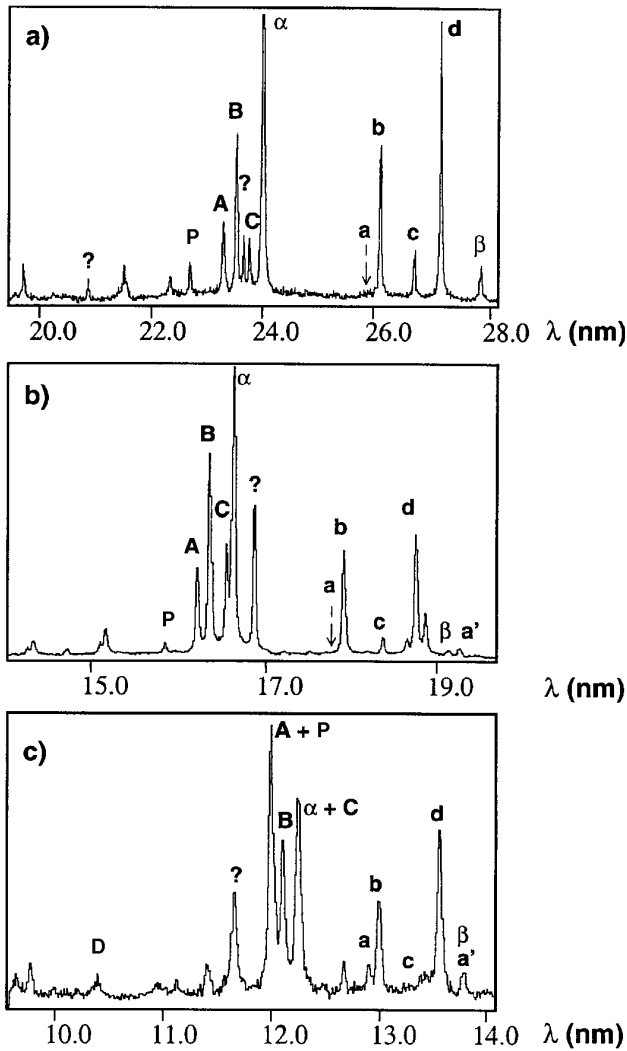}} \caption{The
Boron-like 2p-3s and 2p-3d  sextet electric-dipole transitions for
(a) O IV, (b) F V and (c) Ne VI. P: 1s2s2p$^{3}$ $^{6}$S$^{o}$ - 1s2s2p$^{2}$3s$%
^{6}$P electric-dipole transitions. D: 1s2s2p$^{3}$ $^{6}$S - 1s2s2p$^{2}$3s$%
^{6}$P electric-dipole transitions. A, B and C: 2s2p$^{2}$
$^{4}$P-2s2p($^{3}$P$^{o}$)3d $^{4}$P$^{o}$, $^{4}$D$^{o}$ and
$^{4}$F$^{o}$. a, b and c: 2s2p$^{2}$
$^{2}$D-2s2p($^{3}$P$^{o}$)3d $^{2}$P$^{o}$, $^{2}$F$^{o}$ and
$^{2}$D$^{o}$. d and a': 2s2p$^{2}$ 4P-2s2p($^{3}$P$^{o}$)3s
$^{4}$P$^{o}$ and 2s2p$^{2}$ $^{2}$S-2s2p($^{3}$P$^{o}$)3d
$^{2}$P$^{o}$. a and ß: 2s$^{2}$2p $^{2}$P$^{o}$-2s$^{2}$3d
$^{2}$D and 2s$^{2}$2p $^{2}$P$^{o}$-2s$^{2}$3d $^{2}$S.}
\label{Fig3}
\end{figure}

\begin{table*}
\caption{\label{tab:table1}The wavelengths (in \AA ) of the sextet
transitions of O IV, F V and Ne VI.}
\begin{ruledtabular}
\begin{tabular}{lcccllll}
  Ions & label & term$_{lo}$ & term$_{up}$ & $\lambda $$_{obs}$(\AA
  )by this work
  & $\lambda$$ _{obs}$(\AA ) & $\lambda$ $_{mchf}$(\AA ) & $\lambda$ $_{mcdf}$(\AA ) \\
  \hline
  O IV & P & 1s2s2p$^{3}$ $^{6}$S & 1s2s2p$^{2}$3s $^{6}$P & 227.13(4) & 228.63 $^{a}$ &
  228.70 $^{b}$ & 230.00 $^{a}$ \\
  F V & P & 1s2s2p$^{3}$ $^{6}$S & 1s2s2p$^{2}$3s $^{6}$P & 158.61(3) & 161.39 $^{a}$ &
  161.46 $^{b}$ & 162.10 $^{a}$ \\
  Ne VI & P & 1s2s2p$^{3}$ $^{6}$S & 1s2s2p$^{2}$3s $^{6}$P & 119.98(2) & 120.04 $^{c}$ &
  120.33 $^{c}$ & 120.20 $^{c}$ \\
  Ne VI & D & 1s2s2p$^{3}$ $^{6}$S & 1s2s2p$^{2}$3d $^{6}$P & 103.99(3) &
  106.236 $^{c}$ & 106.27 $^{c}$ &
  106.26 $^{c}$
\end{tabular}
\end{ruledtabular}
a Blanke~\cite{bl} b Miecznik~\cite{mie} c Lapierre~\cite{lap}
\end{table*}

\begin{figure}[tbp]
\centerline{\includegraphics*[scale=0.76]{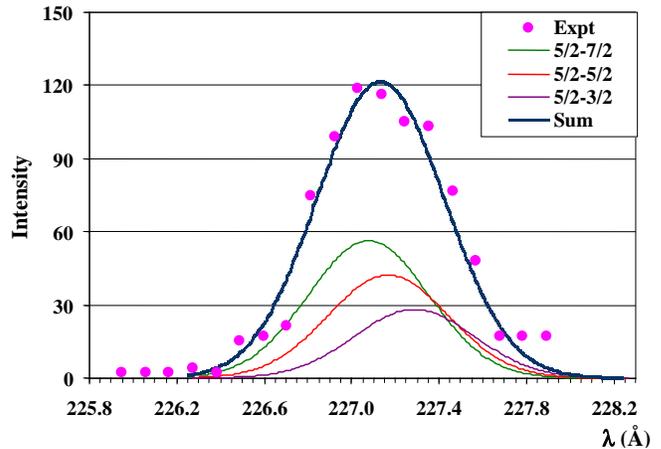}}
\caption{Relative intensity of the 1s2s2p$^{3}$ $^{6}$S$^{o}$ -
1s2s2p$^{2}$3s $^{6}$P transition in O IV. The unit of intensity
is arbitrary. }\label{Fig4}
\end{figure}
The spectra of fluorine recorded in above experiments are shown in
Figs. 2. In the "direct" F$^{5+}$ beam (similar for O $^{4+}$ and
Ne $^{6+}$ ion beams), the fraction of the ions in the
F$^{5+}$(1s$^{2}$2s$^{2}$) state is high, about 97\%, and that of
the F$^{5+}$(1s$^{2}$2s3s) metastable state is low, about 3\%. In
the upper figure of Fig. 2 one sees that the transitions between
doublet terms are dominant. In the "prepared" F$^{5+}$ beam, the
fraction of the ions in the F$^{5+}$(1s$^{2}$2s2p
$^{3}$P$_{0,1,2}$) state increases to about 45\%, that of the
F$^{5+}$(1s$^{2}$2s$^{2}$) state decreases to about 50\%. The
lifetime of the F$^{5+}$(1s2s2p$^{2}$ $^{5}$P$_{1,2,3}$) quintet
metastable state is long enough for such ions to reach the second
excitation region, the cesium cell. The fraction of the ions in
the F$^{5+}$(1s2s2p$^{2}$ $^{5}$P$_{1,2,3}$) quintet metastable
state is about 5\% at the energy ($>$1 keV/amu). The fraction of
transitions between quartet terms increase dramatically, as well
as that of transitions between sextet terms. In the lower figure
of Fig. 2 the transition labelled as P is identified
as the 1s2s2p$^{3}$ $^{6}$S$^{o}$ - 1s2s2p$^{2}$3s$%
^{6}$P electric-dipole transitions, which does not appear in the
direct spectrum. Please be noted that the 2s2p$^{2}$
$^{2}$D-2s2p($^{1}$P$^{o}$)3d $^{2}$F$^{o}$ transition is at
wavelength of 158.95 \AA \ in the direct spectrum, which decreases
dramatically in the prepared spectrum, whereas the 1s2s2p$^{3}$ $^{6}$S$^{o}$ - 1s2s2p$^{2}$3s$%
^{6}$P electric-dipole transition at wavelength of 158.61 \AA \
 shows up.

Shown in Fig.3 are the prepared spectra of oxygen, fluorine and
neon obtained in above beam-gas experiments. The 1s2s2p$^{3}$ $^{6}$S - 1s2s2p$^{2}$3s, 3d $%
^{6}$P electric-dipole transitions in O IV, F V and Ne VI have
been searched in these clean spectra. The observed 1s2s2p$^{3}$ $^{6}$S - 1s2s2p$^{2}$3s $%
^{6}$P electric-dipole transitions at wavelength of 120.04 \AA \
for Ne VI ~\cite{lap} is confirmed. The observed 1s2s2p$^{3}$ $^{6}$S - 1s2s2p$^{2}$3s, 3d $%
^{6}$P electric-dipole transitions at wavelength of 228.63 \AA \
for O IV, 161.39 \AA \ for F V and 106.232 \AA \ for Ne
VI~\cite{bl,lap} are absent. Meanwhile, we have found some notable
features were unidentified, which are unable to be seen in direct
spectra. Shown in Fig. 4 are the details of the
1s2s2p$^{3}$ $^{6}$S$^{o}_{5/2}$ - 1s2s2p$^{2}$3s $%
^{6}$P$_{j}$ transition in O IV. Here the transition rates from
the fine structure j=7/3, 5/2 and 3/2 of the upper state were the
results of single-configuration Hartree-Fock (SCHF) calculation by
this work. The wavelengths of the fine structure components were
calculated SCHF results plus a fitted shift for all three
components. The Gaussian curves of the three components utilized
the experimental width of 0.6 \AA \ for oxygen spectrum. The
summation of the three fine structure components is a
least-squared fitting of the measured data. The measured
wavelength of the transition of 227.13$\pm 0.04 $ is the weighted
center of the fitted profile of experimental data using the above
theoretical analysis. After studying the details of the
transitions theoretically and experimentally described above, and
comparing with the measured results of O IV,F V and Ne VI from
beam-foil experiments~\cite{bl,lap}, and multi-configuration
Hartree-Fock (MCHF) and multi-configuration Dirac-Fock (MCDF)
calculations of F V and Ne VI~\cite{bl,mie,lap},
we assign some unidentified observed lines as the 1s2s2p$^{3}$ $^{6}$S - 1s2s2p$^{2}$3s, 3d $%
^{6}$P electric-dipole transitions in O IV, F V and Ne VI. Results
of the identification and measurements of wavelengths of
transitions between sextet states by this work and comparison are
shown in Table I. Errors of wavelengths are small mainly from
calibration and curve fitting. The latter includes experimental
and statistical errors.

Present results on O IV, F V and Ne VI represent the first
explicit experimental data on transitions between sextet states in
boronlike ions. Well prepared beam-gas spectroscopic experiments
have yielded new and explicit information on the system. Using the
calculated wavelengths and transition rates by this work and
results by others, we were able to assign the observed lines to
the 1s2s2p$^{3}$ $^{6}$S$^{o}$-1s2p$^{2}$3s,3d $^{6}$P
electric-dipole transitions in O IV, F V and Ne VI, and measured
the wavelengths with good accuracy.

These measurements are part of a series aimed at understanding
optical emissions and energy terms in doubly excited sextet states
in boronlike ions. The experiments also lend reliability to the
use of MFHD and MCDF approaches in calculating wavelengths and
transition rates for the transitions between sextet states in
boronlike ions.

We thank J. D\'{e}sesquelles for helpful discussions.



\end{document}